**Wavelength modulation laser-induced fluorescence for plasma characterization**

I. Romadanov[1], Y. Raitses[1], A. Smolyakov[2]

The author to whom correspondence may be addressed: iromada2@pppl.gov

[1]Princeton Plasma Physics Laboratory, Princeton, NJ 08543, USA

[2]Department of Physics and Engineering Physics, University of Saskatchewan, Saskatoon, Saskatchewan S7N 5E2, Canada

**Abstract**

Laser-Induced Fluorescence (LIF) spectroscopy is an essential tool for probing ion and atom velocity distribution functions (VDFs) in complex plasmas. VDFs carry information about kinetic properties of the species critical for plasma characterization, yet their accurate interpretation is challenging due to multicomponent distributions, broadening effects, and background emissions. Our research introduces wavelength modulation (WM) LIF as a solution to enhance VDF sensitivity measurements. WM-LIF, unlike standard amplitude modulation (AM) methods, measures the derivative of the LIF signal, providing greater sensitivity to changes in VDF shape. A numerical model was developed to compare the efficacy of WM and AM signals in lock-in amplifier-based measurements. Experiments were conducted in a weakly collisional argon plasma with magnetized electrons and non-magnetized ions. The argon ion VDFs were measured using a narrow-band tunable diode laser, with its wavelength scanned across the Ar ion transition $4p^4D_{7/2} - 3d^4F_{9/2}$ centered at 664.553 nm (in vacuum). A lock-in amplifier detected the second harmonic WM signal, generated by modulating the laser wavelength with an externally controlled piezo-driven mirror. Our findings indicate that WM-LIF can be used to determine VDF parameters, such as distribution components and their temperatures and velocities, serving as an independent method of obtaining such parameters, which can be used for improving analysis of conventional AM approach. This method is especially beneficial in environments with substantial light noise or background emissions e.g., thermionic cathodes and reflective surfaces.

1. **Introduction**

LIF spectroscopy [1,2] is a diagnostic tool that is used to determine spatially [3,4] and temporally [5,6] resolved measurements of spectral line profiles of ions or atoms under complex plasma conditions. These conditions commonly occur in laboratory and industrial environments, including plasma processing, and electric space propulsion [7–10] applications. In weakly collisional plasmas, the Doppler effect is a primary broadening mechanism of the spectral line profile representing the VDF. This statistical function reveals crucial kinetic properties of plasma, such as temperature and velocities. This paper focuses on LIF measurements in weakly collisional plasmas with non-equilibrium argon ion VDFs (IVDF), excited from metastable levels. Some results are also shown for Doppler-shifted IVDF due to a directed ion flow.

Interpreting VDFs can be ambiguous, especially when dealing with closely located velocity group peaks, partially or entirely overlapped distributions [11–13]. Fluctuations in plasma background emission, measurement noise, and other broadening mechanisms like the Zeeman [14] or Stark effects [15] further complicate an accurate VDF description. Hence, it is critical to establish a robust methodology for VDF measurements to reduce uncertainty in understanding plasma dynamics.

Derivative spectroscopy [16], predominantly used in absorption measurements, can be used for quantifying complex absorption features. This technique focuses on the rate of change in the spectral line shape with respect to wavelength, eliminating broad background absorptions and identifying individual features within complex contours. Derivative spectra can be obtained through post-processing of the raw signal or with a combination of electronic and optical methods like wavelength modulation [17]. Modern diode lasers, with their rapid wavelength, frequency or shift modulation capabilities [18,19], offer an appealing option for acquiring derivative spectra through WM spectroscopy [20].

Typically, LIF measurements utilize laser light amplitude modulation (AM), usually performed with a mechanical chopper, acousto-or electro-optic modulators (AOM or EOM), followed by lock-in detection of the pulsed fluorescence signal. WM spectroscopy, often used for enhanced trace species detection [21], modulates the light's wavelength around a central absorption line at a specific amplitude and frequency. While it is more commonly used for absorption measurements, there have been several studies where fluorescence signals were detected [22–24]. Lock-in detection is used to extract a signal at the $n$th harmonic of the modulation frequency, yielding the derivative of the spectral line profile. WM spectroscopy also shifts signals to a higher frequency region, enabling higher frequency modulation compared to AM, reducing 1/f noise, and enhancing the signal-to-noise ratio [20].

It is worth noting, that similar WM techniques, when system response to the applied modulation is measured, were applied for electrostatic probe measurements of electron [25] or ion energy distributions [26]. Thus, this approach is fundamental across various fields and applications.

In this work, we implement a WM approach for LIF measurements of VDFs in plasma using a tunable diode laser, aiming to provide a more robust VDF analysis method. To validate our approach, we conducted modeling studies highlighting the benefits of the derivative approach and performed experiments where AM and WM LIF spectra were collected at three locations within an industrial plasma source (similar to Bernas source [27]) operating with argon. We demonstrated that, in cases of low signal-to-noise ratio and complex VDF profiles, WM LIF provides more reliable fitting, leading to a more accurate identification of plasma dynamics.

This paper is organized as follows: Section 2 introduces derivative spectroscopy and discusses the WM. Section 3 presents the AM and WM models along with their results. The experimental setup is described in Section 4, and the experimental results are detailed in Section 5. The discussion and comparison of experimental results are provided in Section 6. Conclusions are summarized in Section 7

## 2. Review of derivative and WM spectroscopic techniques
### 2.1. Derivative spectroscopy

Introduced in the 1950s [28,29], derivative spectroscopy (DS) is an analytical technique that enhances resolution and sensitivity of spectroscopic measurements across various applications [30]. In this technique, the spectroscopic data, including absorption or emission spectra, are processed to generate a derivative spectrum. This spectrum represents the rate of change of the original spectrum signal as a function of light wavelength, wavenumber, or frequency.

One of the DS's primary advantages is the ability to enhance line shape analysis by better resolving closely spaced spectral features. In conventional spectra, these features often blend, making it challenging to distinguish individual components. DS addresses this by examining the derivatives of these spectral lines' intensities with respect to the wavelength, enabling a clearer identification of spectral features. The first derivative helps identify the location of a peak, denoted by a zero-crossing point. The second derivative pinpoints the areas of highest curvature in the normal spectrum, thereby improving the resolution of closely spaced spectral features, making overlapping peaks more distinguishable, and improving the detectability of subtle spectral changes. Figure 1a shows an example of synthetic Maxwellian VDF ($f$) for species with zero mean velocity and temperature of 0.1 eV. First $f'$ and second $f''$ derivatives are shown as well. One of the benefits is the cancellation of the background offset. Figure 1b provides an example of synthetic VDF signal resulting from two closely located distributions $f_1$ and $f_2$ with parameters as in Fig 1a and mean velocities of +/-300 m/s (converted to corresponding laser light frequency shift). While resulting VDF ($f$) can be misinterpreted as a single Maxwellian VDFs, second derivative makes its true shape immediately obvious. Figure 1c illustrates an example of improving the detectability of subtle spectral changes, when small fast Maxwellian distribution $f_2$ with temperature of 0.05 eV and mean velocity of 800 m/s is completely overlapped by larger $f_1$ Maxwellian distribution with zero mean velocity and temperature of 0.22 eV. Second derivative makes presence of $f_2$ distribution clearly identifiable.

Furthermore, DS effectively handles issues such as line shape skewing, baseline drift, and light scattering in conventional spectra. These disturbances often result from variations in background emission or light scatter from the vacuum vessel, optics, or windows. The derivative spectra allow significant cancelation of these effects, improving description of observed VDFs e.g., Fig. 1d, where a nonlinear background is added to the Maxwellian distribution with parameters as in Fig. 1a. It is important to note that if these effects exhibit a strong nonlinear dependency on wavelength, their cancellation could still result in artifacts. For instance, the second derivative will not become zero for functions of an order higher than two.

Potential challenges of DS technique such as an increased noise in higher-order derivatives and the need for precise measurements to avoid wavelength reproducibility errors can be mitigated through implementing direct measurement techniques, rather than post-processing of the measured signal. Traditional post-processing, which often involves signal smoothing or fitting, can introduce artifacts into obtained derivatives. By directly measuring VDF derivatives using techniques like WM, noise can be effectively suppressed due to the capabilities of lock-in amplifiers. Employing stepwise changes in laser wavelength, rather than scanning, enhances the precision of wavelength measurements. The first and second derivatives typically provide a balance between noise levels and resolution enhancement, highlighting DS's utility in line shape analysis. Further details on this technique can be found in [16,26].

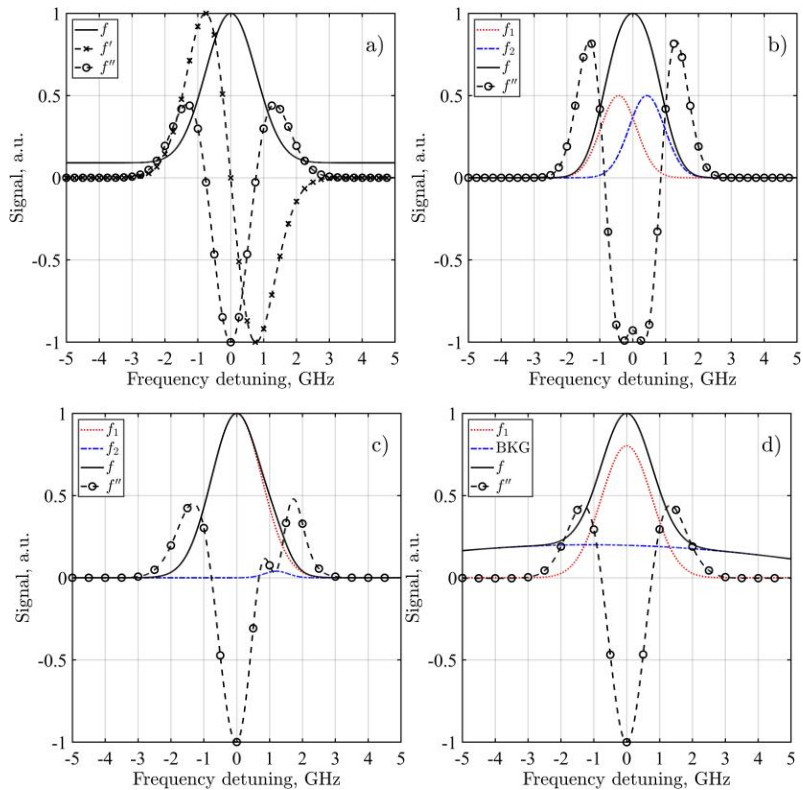

FIG. 1. Illustration of derivative spectroscopy. (a) Single Maxwellian VDF signal (solid black line) and its first ($f'$, dashed line with crosses) and second ($f''$, dashed line with circles) derivatives. (b) Bi-Maxwellian VDF signal (dotted red and dashed blue lines) with nonzero mean velocities and its second derivative $f''$ (dashed line with circles). (c) Bi-Maxwellian VDF signal with a bulk distribution (dotted red) and a smaller faster group (dashed blue) and its second derivative $f''$ (dashed line with circles). (d) Single Maxwellian VDF (solid black line) signal with a nonlinear (sinusoidal) background (BKG, dashed blue line) and its second derivative $f''$ (dashed line with circles). Distributions on each subplot are normalized to the maximum of the total distribution (solid black line). Derivative signal is maximized to the maximum absolute value of the derivative signal.

### 2.2. WM spectroscopy

WM spectroscopy offers high sensitivity and robustness against background noise, making it suitable for challenging environments characterized by strong turbulence or high pressure and temperature [32–36]. While the principles of WM spectroscopy have been extensively covered in the literature [17,20], this paper provides just a basic overview, outlined below and schematically illustrated in Fig. 2. The WM technique comprises the following three elements:

1. Wavelength control consists of the scanning and modulation of light's wavelength around a specified center wavelength at a frequency $\omega_m$ with a designated modulation amplitude $\Delta$ (often referred to as the modulation depth). The capacity to adjust both the modulation amplitude and frequency provides significant flexibility in the measurement process. Typically, modulations are achieved via diode current control, leading to concurrent amplitude modulation. This complicates the analysis of the WM signal, necessitating thorough laser characterization and signal modeling. However, modern diode lasers, which allow for fast (in kHz range) voltage modulation of grating mount piezo actuator, allow for modulation of the laser wavelength with minimal impact on its amplitude. This facilitates data analysis focused solely on wavelength modulation.
2. The laser light is directed through a test sample, in this instance, plasma. Depending on the measurement type, either light absorption or emitted fluorescence is measured. A suitable detector, such as a photodiode, is employed based on the specific scenario.
3. The detector signal is fed to a lock-in amplifier in order to extract a certain harmonic of the detector signal at a detection frequency $n\omega_m$ where $n = 1,2$, etc. with a bandwidth given by the inverse of the integration time (lock-in amplifier time constant). When the frequency modulation amplitude $\Delta$ is much smaller than the width of the absorption profile, the retrieved signal is proportional to the $n$th harmonic of the cosine series of the absorption profile. It can be shown (see Eq. B9 in Appendix B of [20]) that $n$th harmonic of the in-phase component of the lock-in amplifier output is proportional to the $n$th derivative of the lineshape profile. This is the reason why the WM technique is a DS method.

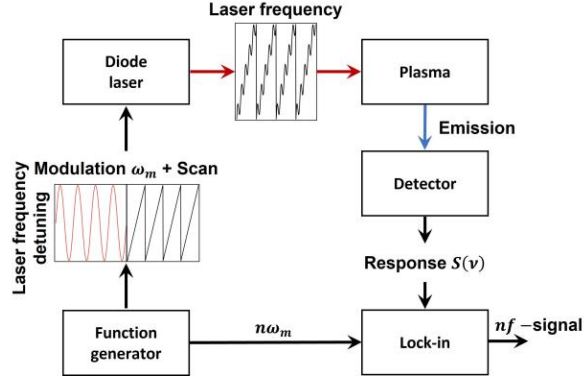

FIG. 2. Schematic representation of WM measurements. The function generator produces overlapping sawtooth and sine waves ($\omega_m$) with distinct frequencies: Hz range (or lower) for the sawtooth and kHz range for the sine. These waves drive the diode laser, generating modulated laser light centered on the probed transition's frequency. This modulated light traverses the sample, and the resulting emission is detected. This detected response is then input to the lock-in amplifier, with its reference frequency set to a certain $n$-th harmonic of the sine signal $\omega_m$.

It's worth mentioning that there's an alternative approach using high-frequency modulation, known as frequency modulation (FM) spectroscopy (see [17,37]). In general, when modulation is performed at arbitrary frequency and with arbitrary amplitude, the rigorous signal modeling and laser behavior characterization become important for a correct recovery of the information from the signal. However, this topic is out of scope of this paper.

### 3. Modeling of WM and AM signals

This section presents an overview of foundation for WM and AM signals modeling, with details being outlined in Appendix A and B. The aim is to illustrate that WM signals enable a more reliable extraction of information from measured LIF signals as compared to AM signals, particularly in high-noise environments. Parameters such as the temperature and flow velocities of constituent atomic or ionic groups are typically determined through a fitting procedure applied to experimental data. However, it is crucial to make initial assumptions about the VDF shape, including the expected number of distributions, mean velocities, and temperatures, to prevent data overfitting. The computation of higher moments of measured distributions (e.g., kurtosis or skewness) [38–40] or other statistical techniques [12,13] can aid in forming these initial assumptions. It is important to note, however, that this analysis can only be performed during post-processing and is significantly influenced by the signal noise and the smoothing or fitting algorithms used for data processing. As demonstrated in the previous section, derivatives of distributions can provide similar insights into the VDF shape. WM allows for direct measurements of VDF derivatives, enabling the fitting of the resulting data without additional post-processing.

The model and data processing presented here emulate the steps of a typical LIF measurement. Laser light, characterized by a very narrow linewidth, is tuned around the absorption profile, exciting groups of atoms or ions at varying velocities due to the Doppler effect. These excited species then emit fluorescence light, which is measured by a photodetector device, such as a photomultiplier tube (PMT). To differentiate the fluorescence signal from background emissions (originating from plasma, reflecting walls, filaments, etc.), the signal is typically modulated. This modulation enables the use of homodyne detection [41] measurement systems, like lock-in amplifiers [42], known for their ability to extract small amplitude signals from noisy environment. Once the scan across the absorption line is completed, a VDF shape is recovered and subsequently fitted with a function that appropriately describes the assumed distribution. Such function should correctly describe (temperatures, mean velocities) all distributions, forming the distribution function.

When analyzing and fitting experimental data, it is essential to consider all effects that could contribute to absorption line broadening. In low-temperature plasmas, several mechanisms can cause broadening, including Zeeman and Stark effects (due to high magnetic or electric fields), Doppler effect, natural broadening, and hyperfine structures [43]. Doppler broadening, typically the most influential factor, results in a Maxwellian absorption line profile when it originates from the thermal motion of atoms. However, if the medium deviates from thermodynamic equilibrium, the profile may no longer be Maxwellian and can assume various shapes [10]. The hyperfine structure is another crucial factor to consider during profile fitting, with transitions having known hyperfine structures, such as those referenced in [44,45] being preferable.

This study focuses on argon plasma with the most abundant argon isotope under conditions where the Zeeman and Stark effects are negligible. Multimodal distribution, consisting of one or several Maxwellians is assumed. Such distributions are common in plasma devices with crossed electric and magnetic fields, see [46–48]. Under these circumstances, the Doppler-broadened profile, representing VDF, as a function of laser light frequency $v$ can be written analytically as follows [13]

$$f(v) = \sum_{k=1}^{N} \frac{c}{v_0^k} \left( \frac{M_i}{2\pi k_B T_i^k} \right)^{1/2} exp\left( -\frac{M_i c^2}{2 k_B T_i^k} \frac{\left(v - v_0^k\right)^2}{\left(v_0^k\right)^2} \right), \tag{1}$$

where $c$ is the speed of light, $v_0$ is an LIF transition central frequency, $M_i$ is the mass of species, $k_B$ is the Boltzmann constant, $T_i^k$ is the $k$-th distribution temperature, and $v_0^k$ is laser frequency corresponding to the mean velocity of the of the $k$-th distribution in GHz, and $N$ is the total number of distributions present in the plasma.

Various types of lasers, including solid-state lasers, dye lasers, laser diodes, quantum cascade lasers, and optical parametric oscillators, can adjust their wavelength over a broad range. For high-resolution reconstruction of VDF, continuous wave (CW) laser diodes are preferred due to their extremely narrow bandwidth, typically in the MHz range or below. The laser beam intensity profile, as a function of light frequency, is typically represented by a Lorentzian function as follows [20]

$$L(v, v_L) \sim \frac{1}{1 + \frac{(v - v_L)^2}{\Delta \lambda_L^2}}, \tag{2}$$

where $v_L$ is the laser central frequency, which can be tuned, and $\Delta \lambda_L$ laser linewidth. The equation used to establish the relationship between the laser frequency offset and the velocity is as follows

$$\Delta v = -\frac{1}{2\pi} \boldsymbol{v} \cdot \boldsymbol{k}, \tag{3}$$

where $\Delta v$ is the shift in photon frequency from the perspective of the particle, $\boldsymbol{v}$ is the particle velocity vector, and $\boldsymbol{k}$ is the photon wavevector.

Typically, a system is excited at a fixed frequency (sourced from an oscillator or function generator), which is also supplied as an input to the lock-in amplifier. The amplifier then identifies the system's response at this reference frequency. In the context of LIF, a response signal $S(v)$ represents a fluorescence signal, which is proportional to $\int f(v) L(v, v_L) \, dv$, where the laser central frequency $v_L$ is scanned across the absorption line. When this signal is measured across a range of light frequencies from $v_1$ to $v_2$, the output signal from the two-phase lock-in can be expressed as follows

$$X = \int_{\nu_1}^{\nu_2} S(v)\sin(2\pi\omega_{ref}v + \phi_{ref})dv,$$

$$Y = \int_{\nu_1}^{\nu_2} S(v)\cos(2\pi\omega_{ref}v + \phi_{ref})dv, \qquad (4)$$

where $\omega_{ref}$ and $\phi_{ref}$ represent the frequency and phase of the reference signal, respectively, $X$ and $Y$ quantities represent the signal as a vector relative to the lock-in reference oscillator. The $X$ variable is called the 'in-phase' component and $Y$ the 'quadrature' component, for more details see Ref [36]. By calculating the magnitude ($R$) of the signal vector as $\sqrt{X^2 + Y^2}$, the phase dependency is eliminated.

For the model simplicity, it is assumed that $\phi_{ref}$ aligns with the phase of the response signal and can thus be omitted. The sweeping range is divided into multiple intervals to recover the Doppler-broadened absorption line (representative of a VDF profile) or its derivatives. The distinction between AM and WM signals is due to different methods of producing the $S(v)$ signal. In the case of AM, $L(v, v_L)$ is a pulse wave function, with amplitude changing between 0 to $I$. Conversely, in the WM scenario, the amplitude remains constant or oscillates around a certain level, but $v_L$ varies according to a function described by Eq. B1. The specifics of both methods, examples of signal shapes, and laser responses to the modulation are described in the Appendixes A and B.

Modeling results, presented below were obtained by numerically integrating the above equations for AM and WM cases for the following set of parameters, which are relevant to those observed in the experiments with the studied ion source. A bi-Maxwellian singly charged argon ions distribution was assumed, with a bulk distribution at $T_i^1 = 0.17$ eV and zero mean velocity, and a colder, faster distribution with $T_i^2 = 0.04$ eV and a mean velocity of 0.9 km/s. The ratio of peak densities of the two distributions was set at 0.04. The laser line profile was modeled using Eq. 2 for a linewidth $\Delta\lambda_L = 50$ MHz, which is typical value for laser diodes. A comparison of the laser linewidths and the distribution is provided in Fig. 3. As illustrated, the laser line is significantly narrower than the VDF shape.

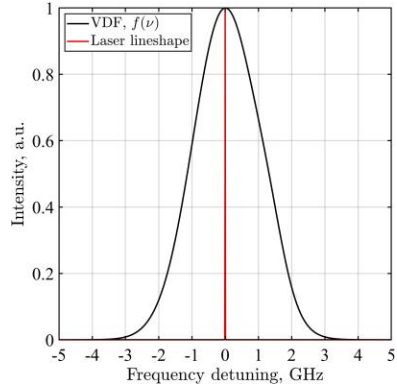

FIG. 3. Laser line shape (red) as compared to Doppler broadened Ar absorption line.

### 3.1. Modeling results

To demonstrate the advantage of the WM technique for the unambiguous extraction of the VDF information from the measured fluorescence signal, Eqs A2 and B2 were numerically solved to obtain lock-in signals with the set of parameters presented earlier. For both cases, the modulation frequency $\omega_m$ was set at 5 kHz, and the lock-in time constant (integration range) was set at 1 second. Modelling was conducted for AM ($S(v, v_{AM})$) and WM ($S(v, v_{WM})$) signals obtained when varying noise levels were added to the $f(v)$ function (see Eq. 1). Noise was modeled as white Gaussian noise with amplitudes ranging from 0 to 0.5 of the maximum signal level.

Examples of the AM signal and WM signal for a noise level equal to half of the LIF signal amplitude are depicted in Figs. 4a and 4b respectively. Both signals are normalized to the maximum amplitude. The analytical signal was obtained by using Eq. 1. The analytical form of the WM signal was obtained as the second derivative of Eq. 1, expressed as follows

$$f''(v) = \frac{2}{\sqrt{\pi}} \sum_{k=1}^{N} \frac{2(v-v_0^k)^2 - a_k^2}{a_k^5} exp\left(-\frac{(v-v_0^k)^2}{a_k^2}\right), \quad (7)$$

where $a_k^2 = \frac{(v_0^k)^2}{c^2} \frac{2 k_B T_i^k}{M_i}$. The background signal is not accounted for in Eq. 1. However, it is worth noting that the form of Eq. 7 remains unchanged in the presence of linear background signal variations, as the second derivative is zero in such cases.

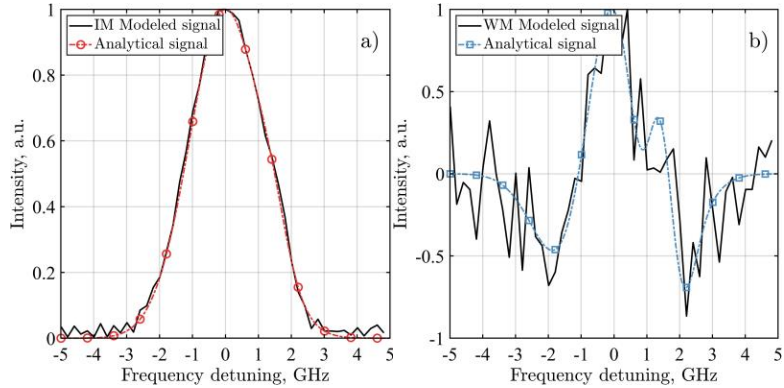

FIG. 4. (a) AM signal from the amplitude modulation model (black solid line) and its analytical form (red dashed line with circles). (b) WM signal from the wavelength modulation model (black solid line) and its analytical form (blue dashed line with squares). Noise level is half the peak of the modeled fluorescence signal.

When experimental data is processed, the obtained signals are fitted with a single distribution or a sum of several distributions. Fitting was performed in MATLAB with weighted nonlinear residuals fit function – "lsqnonlin". The quality of the fit is evaluated using reduced $\chi_{red}^2$, defined as follows:

$$R^2 = 1 - \frac{\Sigma(f_{model} - f_{fit})^2}{\Sigma(f_{model} - \overline{f}_{model})^2}, \quad (8)$$

where $f_{model}$ is the modeled function, and $f_{fit}$ is the fitted curve. This procedure is applied to the signals modeled at three different noise levels (noise amplitudes where 0.1, 0.3, and 0.5 of the signal amplitude) and Eqs 1 and 7 were used as fitting functions, with the number of distributions ($N$) varying from 1 to 5. $R^2$ scores were calculated for each case, and $R^2$ score as a function of $N$ for each noise levels are shown in Fig. 5.

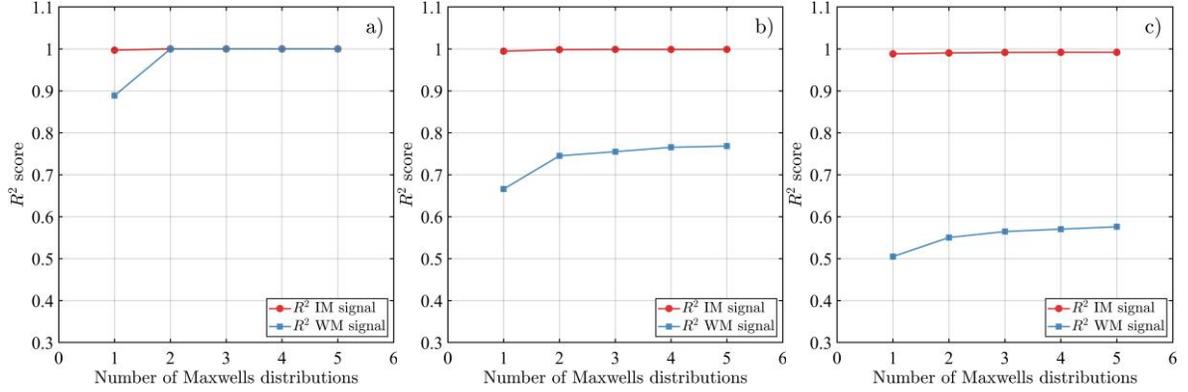

FIG. 5. $R^2$ scores for the fits of modeled AM and WM signals at varying noise levels with varying number of fitted distributions ($N = 1$ to 5). (a) Noise level at 0.1 of maximum signal. (b) Noise level at 0.3 of maximum signal. (c) Noise level at half the peak of the modeled fluorescence signal. Red line with circles represents AM signals; blue line with squares represents WM signals.

The data in Fig. 5 require additional explanation. One notable difference between the fitting of AM and WM signals is the consistently higher $R^2$ value for the AM signal. This can be explained by the lower signal to noise ratio for the WM signal, as can be seen in Fig. 4a and b. Lower WM signal can be explained by comparing Eq. 1 and Eq. 7, which shows that the amplitude of the WM signal is lower than the AM signal by a factor $\left(2(v - v_0^k)^2 - a_k^2\right)/\sqrt{\pi} a_k^4$. Despite lower $R^2$ scores for WM signals, the fitting reliably determines the values of $V$ and $T$, as detailed in Table C1 in the appendix. A closer analysis shows that the $R^2$ value for WM signals is more sensitive to the number of distributions ($N$) used in the fitting. Unlike the AM signal, where the $R^2$ score remains constant across all $N$ values, the WM signal shows a distinct transition between underfitting and overfitting, identifying the correct number of distributions ($N = 2$). Thus, while the AM signal offers lower noise, its fitting results can be ambiguous. In contrast, the WM signal's sensitivity to fitting parameters makes it more suitable for detailed analysis and plasma characterization.

### 4. Experimental setup

Experiments were conducted in the experimental setup described in Ref. [50]. The setup includes a standard 10" diameter six-way cross vacuum chamber. A weakly collisional plasma was generated by a low pressure (0.5 mTorr) argon discharge with a hot thermionic cathode with the applied electric and magnetic fields. In the experiments, the magnetic field was varied between 15-150 Gauss. The discharge voltage was 50-100V. The plasma source features a 1 cm diameter opening on the longer front wall (see Fig. 6). A laser beam, with wavevector $\overline{k}$ and frequency $v$, is launched through an opening on the front wall. The LIF signal is collected through three additional 3 mm diameter holes on the side wall, which are referred as central (on the central line), middle, and edge (closes to the front surface with the opening). The electron temperature, measured using a sweeping Langmuir probe [51], was found to be approximately 5 eV. Given that no other acceleration mechanisms are present in this system, the maximum expected velocity is the Bohm velocity, which is approximately 3.5 km/s. Therefore, it is anticipated that the observed velocities will fall within the range of $\pm 3.5$ km/s.

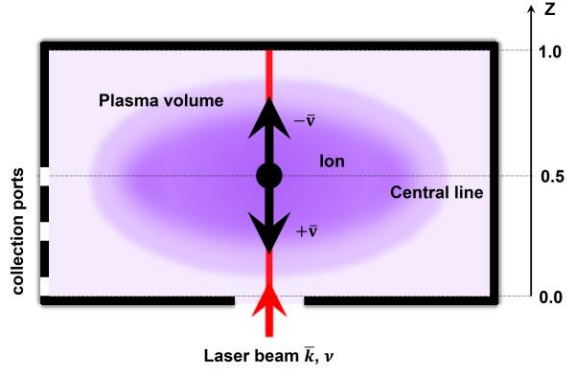

FIG. 6. The schematic of the plasma source illustrates a plasma volume with a diagnostic access opening of less than 1 cm on the front wall, which is not depicted to scale. The path of the laser beam is indicated in red, while the directions of positive and negative velocities are represented by black arrows.

### a. LIF transition

The LIF measurements were conducted by sweeping and simultaneously modulating the frequency of a narrow linewidth, tunable diode laser across the absorption line of an argon ion, which experienced broadening due to the Doppler shift. It was confirmed by previous measurements [50] that the magnetic field used in these experiments does not affect the line broadening. The selected Ar ion transition $3d^4F_{9/2} - 4p^4D_{7/2}$ at 664.553 nm (in vacuum) and fluorescence at 434.929 nm is depicted in Fig. 7.

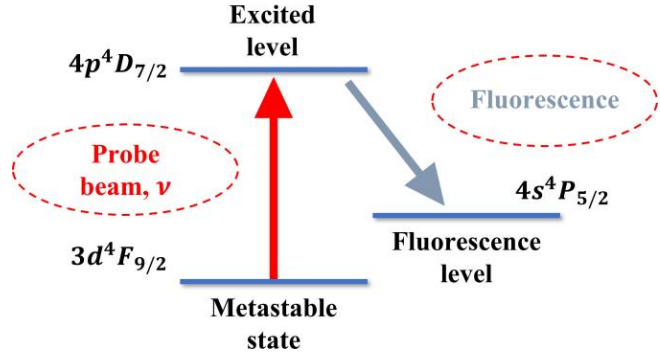

FIG. 7. LIF transition for Ar-II (argon ion).

### b. LIF setup

The LIF system, shown in Fig. 8, is built around a Toptica DLC DL PRO 670 single-mode tunable diode laser (TDL). This Littrow-type grating-stabilized external cavity diode laser offers a coarse tuning range from 660 to 673 nm and a mode-hop free range up to 20 GHz. Depending on the wavelength, the output power reaches a peak of approximately 23 mW. The system maintains a short-term linewidth stability of 600 kHz over 5 μs. The emitted beam, elliptical in shape, is Gaussian with a typical size around 3 mm.

The laser wavelength is controlled by simultaneous scanning and applying a sinusoidal modulation to the voltage directed at the piezo actuator from a signal generator. When modulation is applied to the piezo actuator, the laser power remains constant, avoiding complications related to the residual amplitude modulation effects [52,53]. Scanning was performed using step functions, incrementally increasing the voltage. The modulation frequency was set at $f_m = 1.5$ kHz, and a complete scan across the absorption profile took approximately 150 seconds. This laser model can accommodate modulation frequencies up to

3 kHz. The modulation depth was selected to be about a quarter of the Doppler-broadened spectral line width. The primary constraint on the modulation depth was ensuring the scan remained mode-hop free.

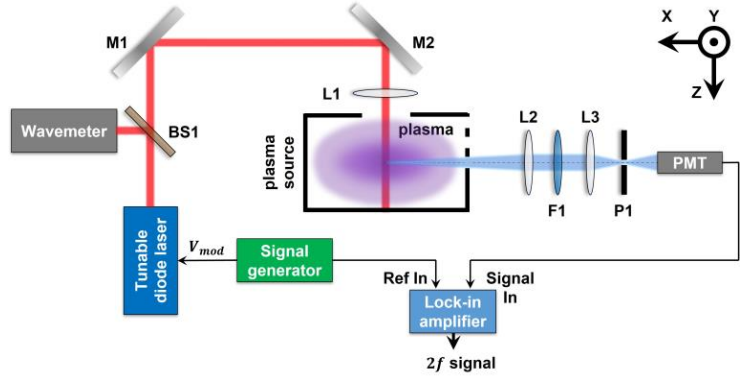

FIG. 8. Block diagram of WM LIF setup and beam path into the plasma source. BS1 – beam splitter; M1,2 – mirrors; L1,2,3 – lenses; P1 – pinhole; F1 – bandpass filter; PMT – Photomultiplier tube.

A more comprehensive description of the setup, along with its schematic, is available in Ref. [50]. The AM modulation setup differed primarily by the inclusion of a mechanical chopper, which was used to modulate the laser intensity amplitude.

## 5. Experimental results and discussions

Measurements were conducted at three locations, each repeated three times. Fig. 9 displays the averaged AM signals, which represent argon IVDF profiles (red line with circles), and WM signals, which represent the second derivatives of IVDFs (blue line with squares), along with standard deviations (shaded area). In these experiments, the true shape of the IVDFs is not known a priori; thus, the obtained IVDFs are first evaluated visually to assess fitting results, presented below. At the center location (Fig. 9a), the IVDF features two distinct peaks, one near 0 km/s and another at -2.5 km/s. The middle location (Fig. 9b) shows a IVDF with a single peak at 0 km/s and asymmetrical tails in both directions, the negative being more pronounced. At the edge location (Fig. 9c), the IVDF displays a complex shape with a central peak at ~0.5 km/s and asymmetrical tails. Based on these observations, it is anticipated that the $R^2$ scores will saturate at $N = 2$ for the center location, and at $N \geq 3$ for the middle and edge locations.

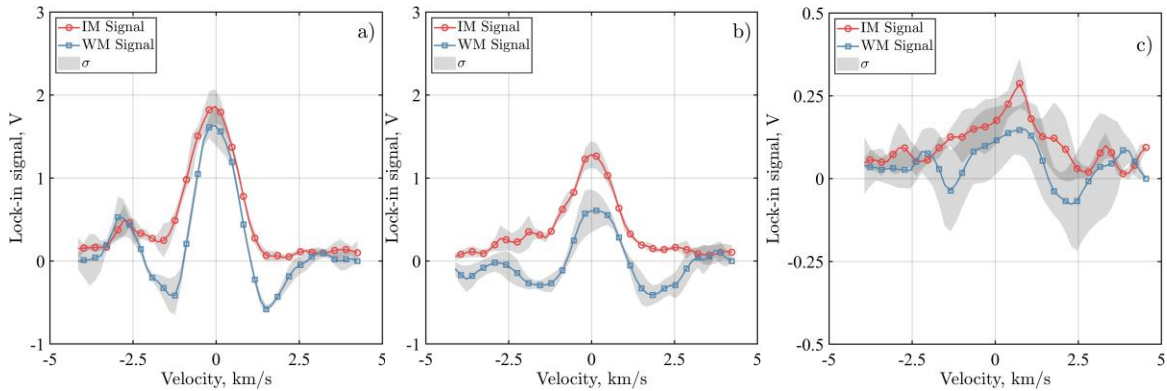

FIG. 9. Experimentally obtained AM signals or IVDF profiles (red, circles) and their WM signals, or IVDF second derivatives (blue, squares) for three locations (central, middle, and edge). The shaded area represents the standard deviation of three measurements.

To evaluate the data, AM and WM curves were fitted at each position using Eq. 1 and 7, respectively. The laser light frequency $v$ was converted to velocity using relationship (3). Number of distributions $N$ was varied from 1 to 5 to emphasize variations in fitting outcomes, and goodness of fit was assessed using the $R^2$ score. The optimal $N$ was identified as the value after which $R^2$ variations are not significant, thus indicating transition to overfitting cases. Fig. 10 displays the corresponding $R^2$ changes, while Figs 11 and 12 present the optimal fits for AM and WM signals, respectively. Further analysis is provided below.

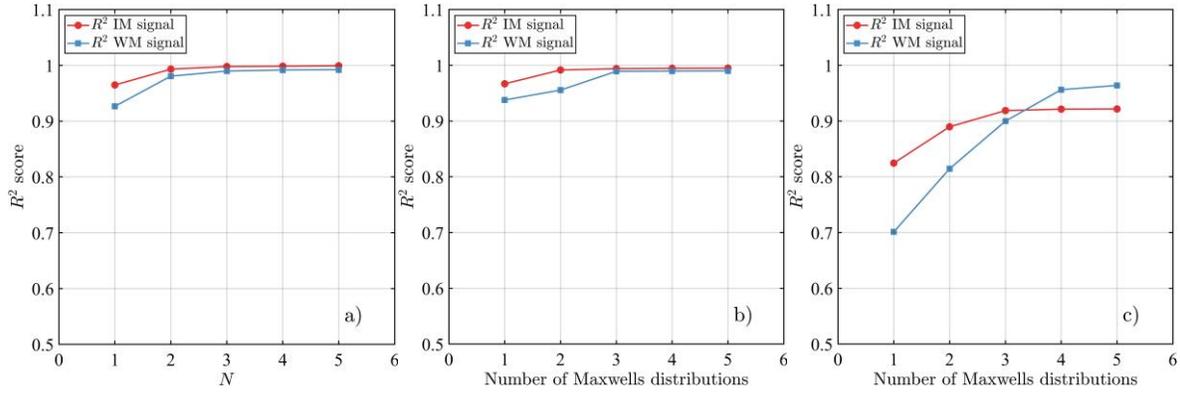

FIG. 10. $R^2$ score evolution of fitting with number of Maxwellian distribution $N$. a) central, b) middle, and c) edge positions.

For the central location measurements (Fig. 10a), the $R^2$ scores for both AM and WM signals exhibit similar trends, reaching maximums after $N = 2$. This is consistent with the initial assessment, which suggested the presence of two distributions. The magnitudes of the $R^2$ values for both signals are also comparable. These characteristics closely resemble those observed in the low-noise modeling case (Fig. 8a), with the only difference that AM fitting of real signal shows greater variation in the underfitting scenario $N = 1$ due to the more pronounced second distribution.

For measurements at the middle location (Fig. 10b), the $R^2$ score for the AM signal behaves similarly to that at the central location, reaching maximum after $N = 2$. For the WM signal, the $R^2$ score plateaus after $N = 3$. Fitting of WM signal aligns with the initial assessment, which suggested the presence of three distributions. The results indicate that the WM signal allows for better capturing of the complex IVDF structure, as this signal is more sensitive to the fitting function.

At the edge location (Fig. 10c), the signal is more susceptible to noise, making evaluation less straightforward. The overall trends for the $R^2$ score differ from previous cases. For AM signal it has a pronounced transition from underfitting case ($N = 1,2$) to the optimal fit cases, reaching a maximum after $N = 3$. For the WM signal, the $R^2$ score consistently increases and plateaus after $N = 4$. Notably, the $R^2$ score for the WM case is higher than for the AM case. Upon examining the fitted curves for both AM and WM signals (Fig. 11c and Fig. 12c), it is evident that the WM signal fitting captures more details within the region of expected velocities ($\pm 3.5$ km/s), defined by Bohm velocity. This shows that in case of strong noise WM signal is better for identifying the true distribution shape.

Consequently, the optimal fits are identified as $N = 2$ for the center location, $N = 3$ for the middle location, and $N = 4$ for the edge location. These optimal fits are illustrated in Figs. 11 and 12, with corresponding $N$ and $R^2$ values indicated. A comparative analysis of the mean velocities and temperatures for all three positions is presented in Table 2 in the Appendix D. The velocity and temperature values from both AM and WM fits align within error margins when the determined number of distribution components is consistent between methods. If the number of components differs, the WM method yields more refined values.

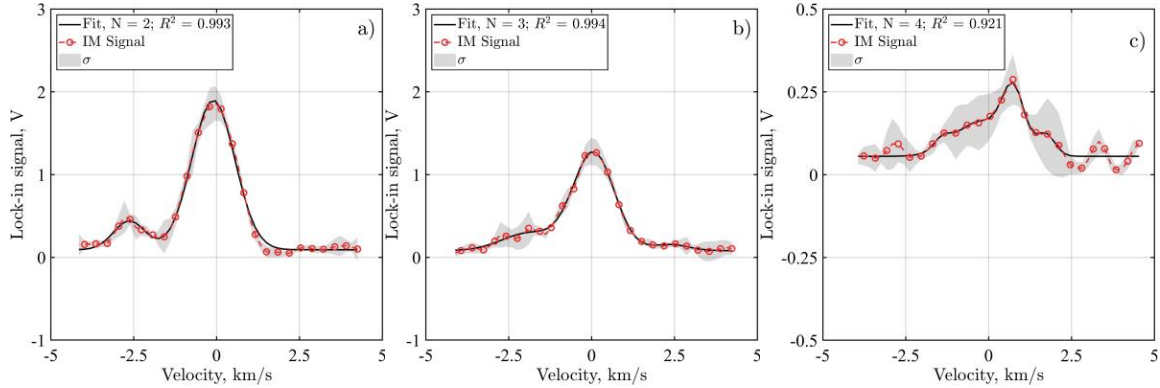

FIG. 11. The best fits of the AM signals for a) central, b) middle, and c) edge positions. $N$ shows the number of Maxwellian distribution.

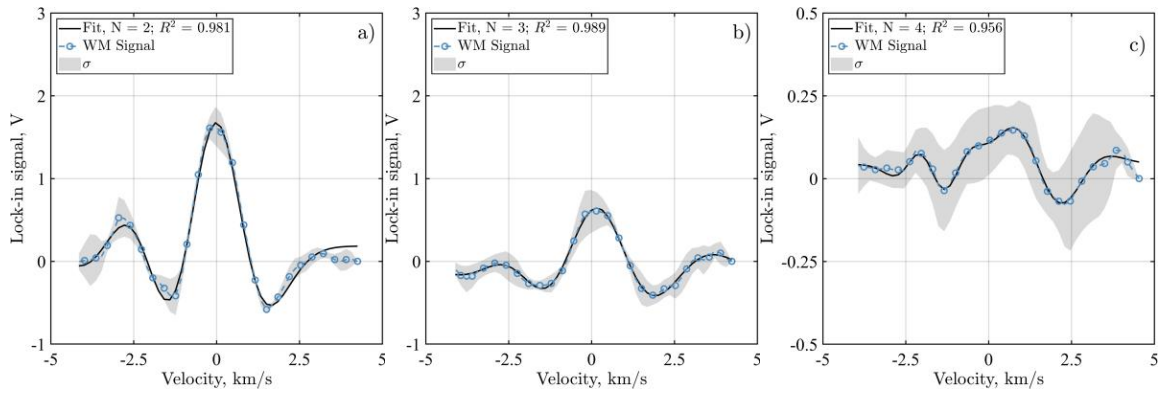

FIG. 12. The best fits of the WM signals for a) central, b) middle, and c) edge positions. $N$ shows the number of Maxwellian distribution used to obtain the fitting curve.

Based on the experimental data it is clear that the complexity of the IVDF shape poses challenges for signal analysis, especially when noise and background variations are significant. The fitting process, guided by the $R^2$ score, shows that $R^2$ dependency on $N$ for modeling and experimental results shows similar trends at different noise levels. This is particularly evident when subtle spectral features are present (see Fig. 9b) or in the conditions of strong noise and weak signal (see Fig. 9c). In all instances, the WM signal fitting is more sensitive to fitting parameters, such as the number of distributions, leading to more reliable data interpretation. The similarity in distribution temperatures and mean velocities obtained from both AM and WM methods further validates the employed methodology.

It is essential to highlight the limitations and drawbacks of the WM technique. First, modulation parameters like amplitude $\Delta$ and frequency $\omega_m$ (see Appendix B, Eq. B1) must be carefully selected. Arbitrary choices for these parameters can distort the resulting signal from the spectral line's derivative [17,31], necessitating a rigorous signal modeling and follow up fitting process. Second, when current modulation is used for wavelength modulation, the presence of residual amplitude modulation (RAM) [54] introduces nonlinear complexities to the signal. This demands either additional laser characterization [40], more sophisticated signal fitting models [41], or methods to mitigate this effect [42]. The reduced WM signal amplitude, as compared to AM signal, may necessitate longer acquisition times to achieve an acceptable signal-to-noise ratio. Lastly, the WM-LIF approach requires careful consideration of modulation parameters and signal interpretation. However, even when direct AM measurements are more straightforward, WM-LIF can serve as a valuable complementary technique for verifying results.

## 6. Conclusion

In this paper, we explored the application of WM spectroscopy to enhance the sensitivity analysis of VDFs obtained through LIF measurements. While WM-LIF measurements have been conducted before, this approach has not been previously applied to VDF analysis and interpretation in plasmas. Our modeling showed that the WM signal allows for a more accurate identification of the true VDF shape, benefiting from the increased sensitivity of the DS technique to shape variations. Analytically it comes from the fact that, contrary to AM signal, the amplitude of the WM signal depends on the temperature, density, and mean velocity of the probed distribution (see Eq. 7), and it exhibits zero background due to its derivative nature. This sensitivity was confirmed through the fitting of modeled AM and WM signals. Using the $R^2$ score as a metric, the WM signal showed larger variability, in contrast to the AM LIF-derived VDF shape, which can yield misleadingly high $R^2$ values even when underfitted (see Fig. 5).

Experimental measurements in argon plasma, generated by a discharge with a thermionic cathode and applied electric and magnetic fields, confirmed these conclusions. The argon IVDFs and their second derivatives (IN and WM signals respectively) were examined at three distinct locations. The IVDF shapes were characterized using a fitting process that varied the number of fitted distributions or their derivatives, as described by Eqs 1 and 7. Goodness of fit was assessed using the $R^2$ score. In conditions of strong signals (center position), both AM and WM methods produced comparable results, offering no distinct advantage for the WM signal. However, in scenarios with higher noise (middle or edge positions), the WM signal more effectively identified the VDF shape. Both methods yielded similar VDF parameters, such as mean velocities and ion temperatures.

Thus, when applied with the appropriate modulation parameters (frequency and amplitude), WM-LIF can serve either as a replacement for traditional IM-LIF measurements or as a complementary method that offers enhanced sensitivity.

## 7. ACKNOWLEDGEMENTS

This work was performed under the U.S. Department of Energy through contract DE-AC02-09CH11466. The authors would like to acknowledge Nirbhav Chopra for fruitful discussions and comments on the text.

## 8. APPENDIX
### A. AM signal model

The AM signal, yielding the zero-order derivative of the VDF, is obtained when the laser intensity amplitude is modulated through variations in the laser diode current, a mechanical chopper, or an acousto-optic modulator. Concurrently, the laser's central wavelength is scanned across the absorption line. The laser response can be represented as follows

$$v_{AM}(v) = sgn\big(A \cdot \sin(2\pi\omega_m(v - v_0)) + A\big) \cdot (v - v_0), \tag{A1}$$

Where $sgn$ is the sign function, returning $1, 0$ or $-1$ depending on the sign of the input function $f$, $A$ is the oscillation amplitude, $\omega_m = \omega_{ref}$ is the modulation frequency, and $\nu_0$ is the central frequency. The laser frequency and intensity amplitude responses are illustrated in Fig. 13a and b. For the illustration purposes, artificially low $\omega_m$ was selected. In Fig 13a interruptions in line represent cases when amplitude is zero. In this case, the fluorescence signal, detected by photodetector, can be written as (by using Eq. 1 and 2)

$$S(v, \nu_{AM}) = \int f(v) L(v, \nu_{AM}) dv.$$

The shape of $S(v, \nu_{AM})$ function is illustrated in Fig. 13

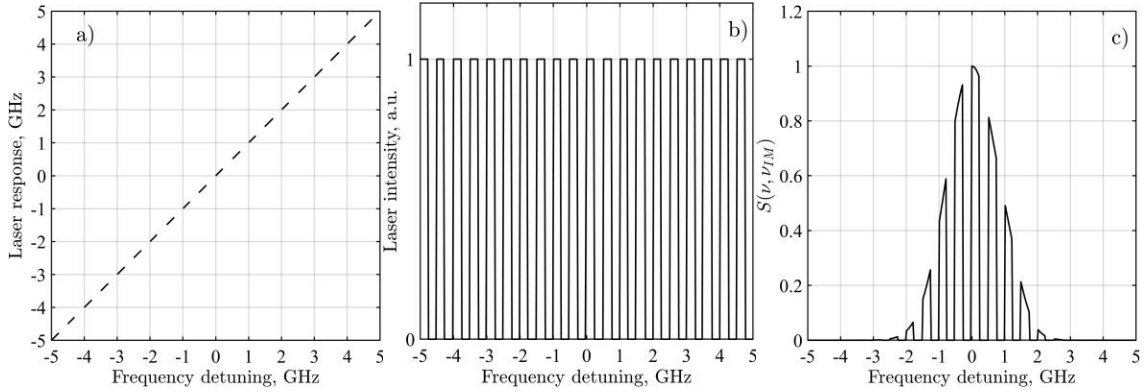

FIG. 13. a) laser frequency response and b) laser intensity response in intensity amplitude modulation configuration. Line interruptions represent zero amplitude of laser intensity. c) $S(v, \nu_{AM})$ function.

The lock-in amplifier signal outputs, from Eq. 4 are written as

$$X_{AM} = \int_{\nu_1}^{\nu_2} S(v, \nu_{AM}) \sin(2\pi \omega_m v) dv,$$

$$Y_{AM} = \int_{\nu_1}^{\nu_2} S(v, \nu_{AM}) \cos(2\pi \omega_m v) dv, \qquad (A2)$$

here, integration is performed across the range of laser scanning frequencies, and this range is defined by the lock-in amplifier constant. The VDF, which is proportional to $S(v, \nu_{AM})$ can be obtained as a magnitude of the signal vector $R = \sqrt{X_{AM}^2 + Y_{AM}^2}$.

### B. WM signal model

Laser light modulation signal was modeled as a combination of a laser light frequency being scanned (linearly) across the absorption line and simultaneously being sinusoidally modulated at frequency $2\pi \omega_m$. Laser light intensity amplitude was assumed to be unaffected by modulations, see details about laser modulation in Section 3. According with the definition of WM, the oscillation frequency is chosen to be $\omega_m \ll \nu_0$, and amplitude $\Delta$, which was chosen to be less than FWHM of the Doppler shifted profile [20]. To summarize all the above, the laser response in case of the WM signal is written as

$$\nu_{WM}(v) = av - \nu_0 + \Delta \sin(2\pi \omega_m (v - \nu_0)), \qquad (B1)$$

where $a$ defines the speed of the sweeping, $\nu_0$ is the central frequency, $\Delta$ is the modulation amplitude, and $\omega_m$ is the modulation frequency. Laser response is illustrated in Fig. 14a.

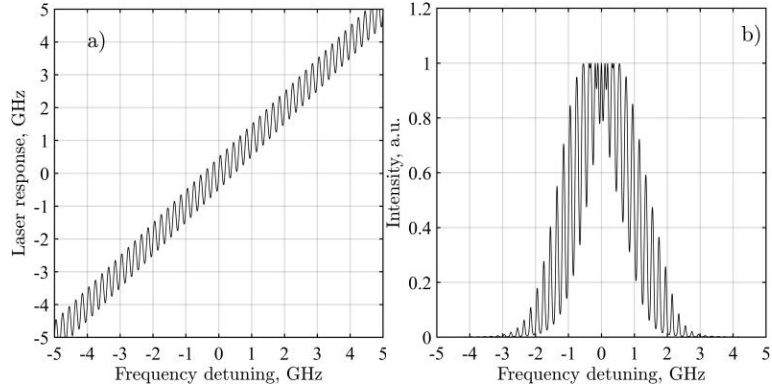

FIG. 14. (a) Laser response in WM configuration; (b) Simulated WM signal $S(v, v_{WM})$, for a scan across of an absorption line.

Like the AM case, the signal, detected by photodetector in this case can be written as (from Eq. 1, 2)

$$S(v, v_{WM}) = \int f(v) L(v, v_{WM}) dv.$$

Note that an explicit analytical expression for this function is not feasible due to the integral containing a product of a Gaussian and a Lorentzian function, where the laser's central frequency in the Lorentzian function is also modulated. Therefore, we solve this equation numerically. An example of the $S(v, v_{WM})$ function is depicted in Fig. 14b, representing the convolution of the VDF shape with the laser response function.

The lock-in amplifier signal outputs (Eq. 4) are written as

$$X_{WM} = \int_{v_1}^{v_2} S(v, v_{WM}) sin(2\pi(2 \cdot \omega_m)v) dv,$$

$$Y_{WM} = \int_{v_1}^{v_2} S(v, v_{WM}) cos(2\pi(2 \cdot \omega_m)v) dv, \tag{B2}$$

here, integration is performed across the range of laser scanning frequencies, and this range is defined by the lock-in amplifier constant. Note that lock-in frequency is set to $2\pi(2 \cdot \omega_m)$, which allows for extraction of the second derivative of the VDF shape. Similarly, to AM case, it is possible to use a magnitude of the signal vector $R$, however, as it was shown in Ref. [20], Eq. B9, the $X_{WM}$ component is proportional to $n$th derivative

### C. Modeling of AM and WM signals with four component distribution

To investigate the $R^2$ score behavior for complex signals with multi-component distributions (illustrated in Fig. 11c), a synthetic signal with $N = 4$, with arbitrarily selected temperatures and velocities, was modeled. This signal underwent the same fitting procedure as applied to experimentally obtained signals. The signals were modeled for both the AM and WM cases, with results presented in Figs. 15a and 15b, respectively. The evolution of the $R^2$ score for fitting both signals as a function of $N$ is depicted in Fig. 15c. It is observed that the $R^2$ score behavior for the AM and WM cases resembles that of the experimental signal, as demonstrated in Fig. 10c.

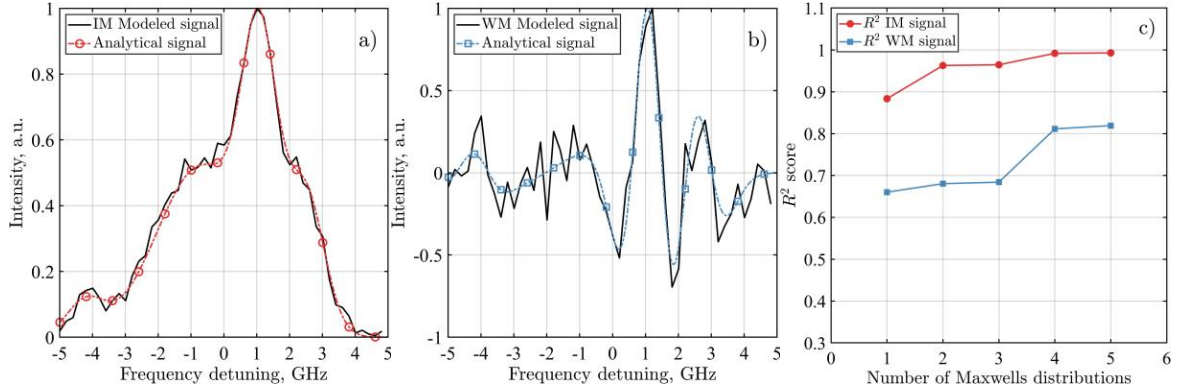

FIG. 15. (a) AM signal from the amplitude modulation model (black solid line) and its analytical form (red dashed line with circles). (b) WM signal from the wavelength modulation model (black solid line) and its analytical form (blue dashed line with squares). Noise level is half the peak of the modeled fluorescence signal.

### D. Comparison of velocities and temperatures for AM and WM fittings

Table 1. Comparison of true and fitted (for N = 2) values of mean velocity and temperatures for modeled signal, shown in Fig. 5. Ground truth values are $V_1 = 0$ m/s, $V_2 = 900$ m/s, $T_1 = 0.17$ eV, $T_2 = 0.04$ eV.

| Noise level | Fit type | $V_1$, m/s | $T_1$, eV | $V_2$, m/s | $T_2$, eV |
|---|---|---|---|---|---|
| 0.1 | AM, N = 2 | 8.0±80 | 0.18±0.03 | 906±80 | 0.04±0.02 |
|  | WM, N = 2 | -0.5±80 | 0.18±0.03 | 899±80 | 0.05±0.02 |
| 0.3 | AM, N = 2 | 11.0±80 | 0.16±0.03 | 941±80 | 0.04±0.02 |
|  | WM, N = 2 | 6.1±80 | 0.18±0.03 | 941±80 | 0.04±0.02 |
| 0.5 | AM, N = 2 | -40.0±80 | 0.14±0.03 | 788±80 | 0.04±0.02 |
|  | WM, N = 2 | -18.0±80 | 0.16±0.03 | 796±80 | 0.04±0.02 |

Error bars for ion temperature is defined as a standard deviation between three measurements. For the ion velocity, the error bar is a combination of standard deviation of three measurements and the uncertainty in laser frequency, converted into corresponding Doppler shift, measurements due to a wavemeter (~60 m/s).

Table 2. Comparison of the values of the mean velocity and temperatures for AM and WM signals obtained experimentally. Errorbars are defined as a confidence interval of the fit (95% interval).

| Location | Fit type | $V_1$, m/s | $T_1$, eV | $V_2$, km/s | $T_2$, eV | $V_3$, km/s | $T_3$, eV | $V_4$, m/s | $T_4$, eV |
|---|---|---|---|---|---|---|---|---|---|
| Center | AM, N=2 | -0.1±0.1 | 0.2±0.01 | -2.6±0.2 | 0.1±0.02 | - | - | - | - |
|  | WM, N=2 | -0.1±0.1 | 0.2±0.01 | -2.5±0.2 | 0.15±0.02 | - | - | - | - |
| Middle | AM, N=3 | 0.1±0.1 | 0.16±0.02 | -1.9±0.5 | 0.28±0.13 | 2.3±0.5 | 0.13±0.27 | - | - |
| | WM, N=3 | 0.1±0.1 | 0.22 ±0.01 | -1.5±0.2 | 0.17 ±0.03 | 2.1±0.2 | 0.30 ±0.07 | - | - |
| Edge | AM, N=4 | 0.8±0.2 | 0.05±0.1 | -0.2±0.9 | 0.25±0.1 | 1.7±0.5 | 0.03±0.1 | -1.4±0.7 | 0.02 ±0.05 |
| | WM, N=4 | 0.5±0.3 | 0.2±0.1 | -0.5±0.3 | 0.2±0.1 | 1.7±0.2 | 0.2±0.1 | -1.6±0.4 | 0.07 ±0.03 |